\documentclass[12pt]{article}
\usepackage{amsfonts}
\usepackage{graphicx}
\usepackage{amssymb,amsmath}

\input{tcilatex}
\begin{document}

\begin{center}
{\LARGE The Gel'fand-Levitan-Krein method and the globally convergent method
for experimental data} \\[5mm]
Andrey L. Karchevsky$^{\bullet }$, Michael V. Klibanov$^{\ast }$, Lam Nguyen$%
^{\bigtriangleup },$ \\[0pt]
Natee Pantong$^{\ast \ast }$ and Anders Sullivan$^{\bigtriangleup }$ \\[5mm]
{\small $^{\bullet }$ Sobolev Mathematical Institute, 4 Acad. Koptyug Ave,
Novosibirsk, 630090, Russia, karchevs@math.nsc.ru\\[0pt]
$^{\ast }$ Department of Mathematics and Statistics, University of North
Carolina at Charlotte, Charlotte, NC 28223, USA, mklibanv@uncc.edu\\[0pt]
$^{\bigtriangleup }$US Army Research Laboratory, 2800 Powder Mill Road
Adelphy, MD 20783-1197, USA,\emph{\ }lam.h.nguyen2.civ@mail.mil and
anders.j.sullivan.civ{@}mail.mil \\[0pt]
$^{\ast \ast }$ Department of Mathematics and Computer Science, Royal Thai
Air Force Academy, Bangkok, Thailand, npantong@hotmail.com}\\[5mm]
\end{center}

{\bf Abstract}: Comparison of numerical performances of two methods for coefficient inverse
problems is described.\ The first one is the classical
Gel'fand-Levitan-Krein equation method, and the second one is the recently
developed approximately globally convergent numerical method. This
comparison is performed for both computationally simulated and experimental data.\vspace{5mm}

{\bf2010 AMS subject classification}: 35R25, 35R30\vspace{5mm}

{\bf Keywords}: 1-d Coefficient Inverse Problems, comparison of two methods, calibration factor


\section{Introduction}

\label{sec:1}

The goal of this publication is to compare performances of two numerical
methods listed in the title on both computationally simulated and
experimental data. The reason why we do not compare other methods is a
practical one: such a study would require much more time and effort than the
authors can afford to spend. We work with the time resolved data of
scattering of the electrical wave field and these data are the same as ones
in \cite{KBKSNF}. The data were collected in the field in the cluttered
environment by the Forward Looking Radar of US Army Research Laboratory
(ARL) \cite{Ng}. The goal of this radar is to detect and possibly identify
shallow explosives.

\subsection{Previous publications}

\label{sec:1.1}

In publication \cite{KBKSNF} the data of the current paper were successfully
treated in the most challenging case: the case of blind data. To do this,
the newly developed the so-called approximately globally convergent
numerical method (AGCM) was applied. Since these experimental data are the
one-dimensional ones, then it is possible to use them for a comparison of
numerical performances of the classical Gelfand-Levitan-Krein integral
equation method (GLK) and AGCM. \textquotedblleft Blind" means that first
the data were delivered to the mathematical group by engineers of ARL,
Sullivan and Nguyen, who coauthor both this paper and \cite{KBKSNF}. No
information about targets was given to mathematicians, except their
placements either above or below the ground.\ Next, the mathematical group
made computations by AGCM and delivered results to engineers. Finally,
engineers compared computational results with the truth and informed
mathematicians about their findings, see Table 1 and Conclusion in
subsection 3.5. The most important observation was that in all available
sets of experimental data the values of computed dielectric constants of
targets were well within limits tabulated in \cite{Table1, Table2}. Since
dielectric constants were not measured in experiments, then this answer was
a quite satisfactory one.

GLK was derived in 1950-ies by Gelfand, Levitan and Krein \cite{GL,Lev,Krein}%
. Originally it was derived for the 1-d inverse spectral problem. Later, it
was shown in \cite{Blag} that an inverse problem for the 1-d hyperbolic
equation can be reduced to the GLK. In addition to \cite{Blag}, we also use
here the material of \S 4 of Chapter 1 of the book \cite{Rom}, where the
derivation of \cite{Blag} is reproduced. As to the numerical results for
GLK, we refer to interesting publications \cite%
{AD,AB1,AB2,HN,Mark,KabSSat,KabS1}.

AGCM was developed for Coefficient Inverse Problems (CIPs) for the
hyperbolic equation $c\left( x\right) u_{tt}=\Delta u,x\in \mathbb{R}%
^{n},n=1,2,3$ with single measurement data, see, e.g. papers \cite%
{BK1,BK2,BK3,KFBPS,KBKSNF,Kuzh,KuzhKl} about this method; a summary of
results is given in the book \cite{BK}. \textquotedblleft Single
measurement" means that the data are generated either by a single point
source or by a single direction of the incident plane wave. Thus, single
measurement means the \emph{minimal amount} of available information. In
particular, the performance of AGCM was verified on 3-d experimental data in 
\cite{BK2,BK,KFBPS}. All known numerical methods for $n-$d, $n>1$ CIPs with
single measurement data require a priori knowledge of a point in a small
neighborhood of the exact coefficient. Unlike this, the \emph{key advantage}
of AGCM is that it delivers good approximations for exact solutions of CIPs
without any advanced knowledge about small neighborhoods of those solutions.
It is well known that this goal is an enormously challenging one.

Hence, a rigorous definition of the \textquotedblleft approximate global
convergence" property was introduced in \cite{BK,BK3,KBKSNF,Kuzh}. In simple
terms, it means that first an approximate mathematical model is introduced.
The approximation of AGCM is a mild one: it amounts to the truncation of an
asymptotic series with respect to $1/s,s\rightarrow \infty ,$ where $s>0$ is
the parameter of the Laplace transform with respect to $t$ of the above
hyperbolic PDE. This truncation is done only on the first iteration. Next, a
numerical method is developed within the framework of the resulting
mathematical model.\ Next, a theorem is proved, which claims that this
method delivers a good approximation for the exact solution of that CIP
without any a priori knowledge of a small neighborhood of that solution. The
common perception of global convergence is that one should obtain correct
solution if iterations would start from almost any point. That theorem
claims, however, that a small neighborhood of the exact solution is achieved
if iterations would start not from any point but rather from a function
which can be obtained without any knowledge of that small neighborhood.\
That theorem was confirmed in numerical studies of both computationally
simulated and experimental data. 3-d experimental data were treated in \cite%
{BK2,BK,KFBPS}. In particular, the most difficult blind data case was
considered in \cite{BK,KFBPS,KBKSNF}. We refer to the paper \cite{Nov} and
references cited there for another non-local method for a coefficient
inverse problem.

\subsection{A huge misfit between real and computationally simulated data}

\label{sec:1.2}

It was pointed out in all publications about AGCM for experimental data \cite%
{BK2,BK,KFBPS,KBKSNF} that there is a huge misfit between measured data and
computationally simulated data, in the case of waves propagation in a
non-attenuating medium. This misfit is evident from a simple visual
comparison of simulated and experimental curves, see, e.g. Figures \ref%
{fig_3} and \ref{fig_4}. This discrepancy is the main difficulty for the
numerical treatment of experimental data by any computational method. In
other words, the experimental data are lying far away from the range of the
operator, which needs to be inverted to solve the CIP. It is well known that
this is a compact operator. Hence, its range is very narrow and the
inversion problem is unstable. Therefore, regularization is necessary.
Still, the regularization theory gives recipes only in the case when the
right hand side of the operator equation is not far from the range of that
operator.\ However, in the case of those experimental data, the right hand
side is actually far away from that range.

Therefore, the crucial step of all above cited publications about
experimental data, was the data pre-processing procedure. This procedure has
extracted such a piece of the data from the whole data, which looked
somewhat similar with the computationally simulated data. Still, it was
impossible to estimate the distance between the extracted data and the range
of that compact operator. Furthermore, those data pre-processing procedures
cannot be rigorously justified neither from the Physics nor from the
Mathematics standpoint. They were based on the intuition only. The single
criterion of their success was the accuracy of resulting solutions of
corresponding CIPs. Since accurate results were obtained for the blind real
data in \cite{BK,KFBPS,KBKSNF}, then those data pre-processing procedures
were unbiased. We describe below the data pre-processing procedure. It
consists of two steps: extracting a piece of data from the whole data and
multiplying this piece by a calibration factor.\ This factor is unknown and
needs to be determined numerically.

In section 2 we describe the experimental data we work with as well as the
data pre-processing procedure. In section 3 we briefly describe the 1-d
version of AGCM.\ GLK is described in section 4. In section 5 we compare
performances of these two methods on our experimental data. We discuss
results in section 6.

\section{Experimental Data}

\label{sec:2}

\subsection{Data collection}

\label{sec:2.1}

\begin{figure}[b!]
\centering
\includegraphics[bb= 0 0 610 540, scale=0.3]{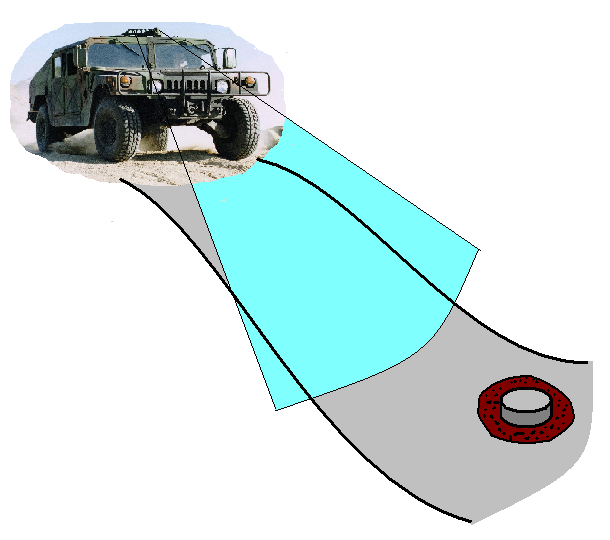}
\caption{\textsf{Schematic geometry of the forward looking radar}}
\label{fig_1}
\end{figure}

\begin{figure}[t!]
\centering
\includegraphics[bb= 0 0 1150 230, scale=0.36]{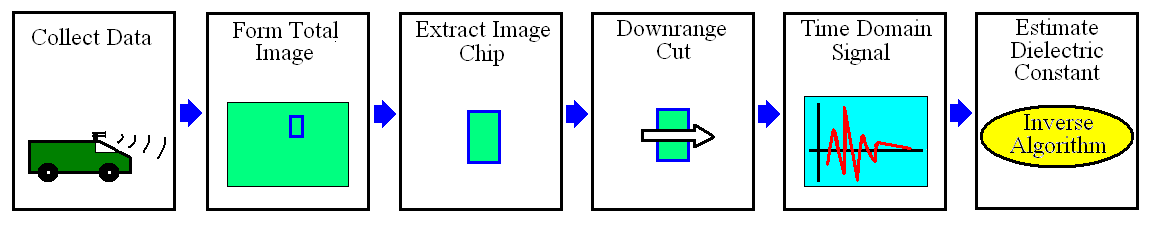}
\caption{\textsf{The schematic diagram of the data collection by the Forward
Looking Radar}}
\label{fig_2}
\end{figure}

The time resolved nanosecond electric pulses are emitted by two sources
installed on the radar (see Fig.~\ref{fig_1}). The schematic diagram of the
data collection by the Forward Looking Radar is depicted on Fig.~\ref{fig_2}%
. Only one component of the electric field is both originated and measured
in the backscattering regime. The data are collected by sixteen (16)
detectors with the time step of 0.133 nanosecond. Only shallow targets
placed either above or a few centimeters below the ground can be detected by
this radar. This is sufficient for antipersonnel plastic land mines as well
as for home made explosives. The depth of the upper part of the surface of a
shallow underground target is a few centimeters. The Ground Positioning
System (GPS) provides the distance between the radar and a point on the
ground located above that target. The error in the latter is a few
centimeters. Time resolved voltages of backreflected signals are integrated
over radar/target distances between 20 meters and 8 meters, and they are
also averaged with respect to both source positions as well as with respect
to readings of sixteen detectors.

Since the radar/target distance is provided by GPS with a good accuracy,
geometrical parameters of targets, including their depths, are not of an
interest here. The main goal of the publication \cite{KBKSNF} was to
calculate ratios $R$ of dielectric constants 
\begin{equation}
R=\frac{\varepsilon _{r}(target)}{\varepsilon _{r}(bckgr)},  \label{1}
\end{equation}%
where $\varepsilon _{r}(target)$ is the dielectric constant of the
background medium. If $\varepsilon _{r}(bckgr)$ is known, then (\ref{1})
enables one to calculate $\varepsilon _{r}(target)$. If a target is located
above the ground, then $\varepsilon _{r}(bckgr)=\varepsilon _{r}(air)=1$.
Unfortunately, dielectric constants were not measured during the data
collection process. Therefore, the authors of \cite{KBKSNF} had no choice
but to rely on tables of dielectric constants \cite{Table1,Table2} when
evaluating the accuracy of results. In our mathematical model (\ref{2.1})-(%
\ref{2.5}) of section 3 $R$ is the function of the spatial variable $x$,
i.e. $R=R\left( x\right) :=\varepsilon _{r}\left( x\right) ,x\in \left(
0,1\right) .$ Hence, after computing $R\left( x\right) $ we took the value $%
\overline{R}$ to compute the dielectric constant of the target $\varepsilon
_{r}(target),$ where%
\begin{equation}
\overline{R}=\left\{ 
\begin{array}{c}
\max_{\lbrack 0,1]}R(x)\ \ \text{if}\ \ R(x)\geq 1,\ \ \forall x\in (0,1), \\%
[2pt] 
\min_{\lbrack 0,1]}R(x)\ \ \text{if}\ \ R(x)<1,\ \ \forall x\in (0,1).%
\end{array}%
\right.  \label{2}
\end{equation}%
Hence, in (\ref{1}) we actually have $R:=\overline{R},$ and by (\ref{2})%
\begin{equation}
\varepsilon _{r}(target)=\overline{R}\varepsilon _{r}(bckgr).  \label{3}
\end{equation}

The importance of estimating dielectric constants of potential explosives is
described in the following citation from the paper \cite{KBKSNF}
\textquotedblleft The recovered dielectric constant by itself is not a
sufficient information to distinguish one target from another. The purpose
of estimating the dielectric constant is to provide one extra piece of
information about the target. Indeed, up to this point, most of the radar
community relies solely on the intensity of the radar image for doing
detection and discrimination of targets. It is hoped therefore that when the
intensity information is coupled with the new dielectric information,
algorithms could then be designed that will ultimately provide better
performance in terms of probability of detection and false alarm
rate\textquotedblright .

\subsection{Severely under-determined data}

\label{sec:2.2}

The experimental data of both \cite{KBKSNF} and this paper are severely
under-determined. We now describe main factors of the under-determination.

\begin{figure}[p!]
\centering
\includegraphics[bb= 0 0 1200 2120, scale=0.26]{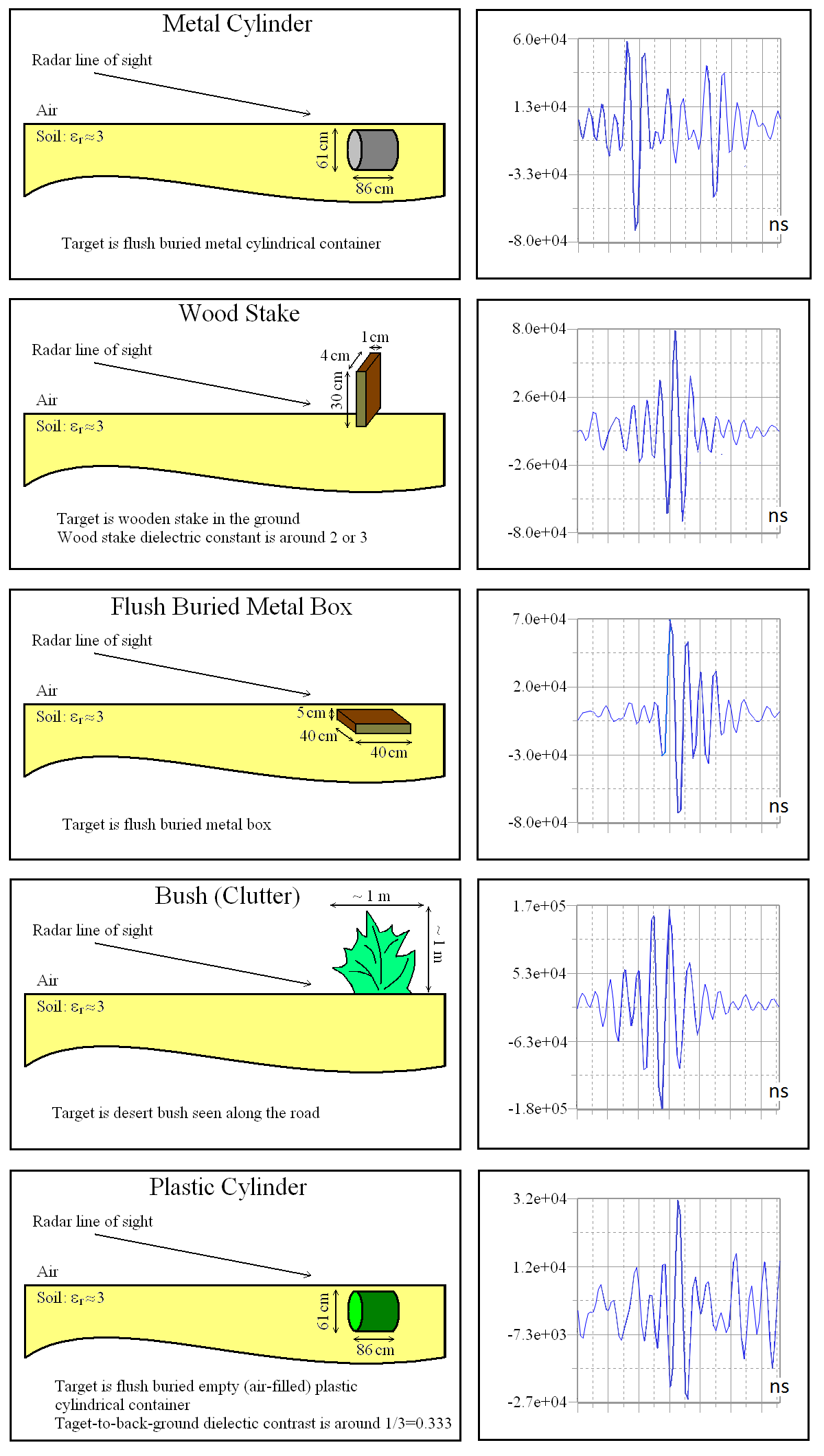}
\caption{\textsf{Targets and collected data before pre-processing}}
\label{fig_3}
\end{figure}

Since targets can be mixtures of constituent materials and since we had
worked with 1-d data, whereas targets are certainly 3-d, then the calculated 
$\varepsilon _{r}(target)$ was a certain weighted average of the spatially
distributed dielectric constant of a target. Due to some technical
limitations, we possess only five sets of experimental data for five
different targets. These targets are schematically depicted on the left
panel of Fig.~\ref{fig_3}. For any target of interest, only a single time
dependent curve can be extracted from the vast amount of data. The right
panel of Fig.~\ref{fig_3} depicts corresponding experimental curves, where
the horizontal axis is time in nanoseconds. An important additional factor
of the under-determination is the clutter surrounding targets. Other factors
contributing to the under-determination were: the integration of all data as
above, the unclear source position in (\ref{2.4}), and it was also unclear
where the time moment $t=0$ is on the data.

\subsection{Data pre-processing}

\label{sec:2.3}

We now describe the data pre-processing procedure. It consists of two stages.

\subsubsection{First stage: selection of a single peak}

\label{sec:2.3.1}

Consider the mathematical model of the above process. Since only one
component of the electric field was both sent into the medium and measured,
and since we have received only one experimentally measured curve per
target, we had no choice but to ignore the complete Maxwell's system and
model the process by only one 1-d wave-like PDE. At the same time, recent
computational testing of the full time dependent Maxwell's system in \cite%
{BM} \ has demonstrated that the component of the electric field $E=\left(
E_{1},E_{2},E_{3}\right) ,$ which was originally sent into the medium,
substantially dominates two other components. In addition, \ we refer to the
accurate performance of AGCM for experimental data in \cite{BK2,BK,KFBPS},
where only a single PDE, which was a 3-d analog of (\ref{2.3}), was used for
the mathematical model.

Let $u(x,t)$, $x\in \mathbb{R}$, $t>0$ be the component of the electric
field which is incident on the medium and which is also measured in the
backscattering regime. Let $\varepsilon _{r}(x)$ be the spatially
distributed dielectric constant of the medium. We assume that 
\begin{eqnarray}
\varepsilon _{r}(x) &\in &[d_{0},d_{1}),\ \ \ \varepsilon _{r}\in C^{1}(%
\mathbb{R}),  \label{2.1} \\
\varepsilon _{r}(x) &=&1,\ \ \ x\notin \left( 0,1\right) ,  \label{2.2}
\end{eqnarray}%
where the numbers $d_{0},d_{1}>0.$ Thus, the interval $(0,1)$ is our domain
of interest in the CIP. We consider this interval as the one whose length is
one meter. We model the process of waves propagation via the following
hyperbolic Cauchy problem 
\begin{eqnarray}
\varepsilon _{r}(x)u_{tt} &=&u_{xx},\ \ x\in \mathbb{R},\ \ t\in (0,\infty ),
\label{2.3} \\
u(x,0) &=&0,\ \ u_{t}(x,0)=\delta (x-x_{0}),  \label{2.4} \\
x_{0} &=&-1.  \label{2.5}
\end{eqnarray}%
Since the source position $x_{0}$ is unclear in our experiment (subsection
2.2), we choose $x_{0}=-1$ in (\ref{2.5}) for the sake of definiteness only.

\textbf{Coefficient Inverse Problem 1 (CIP1)}. \emph{Determine the
coefficient} $\varepsilon _{r}(x)$, \emph{assuming that the following
function} $g(t)$ \emph{is known} 
\begin{equation}
u(0,t)=g(t),\ \ t\in(0,\infty).  \label{2.6}
\end{equation}

\begin{figure}[tb!]
\centering
\includegraphics[bb= 0 0 1400 610, scale=0.3]{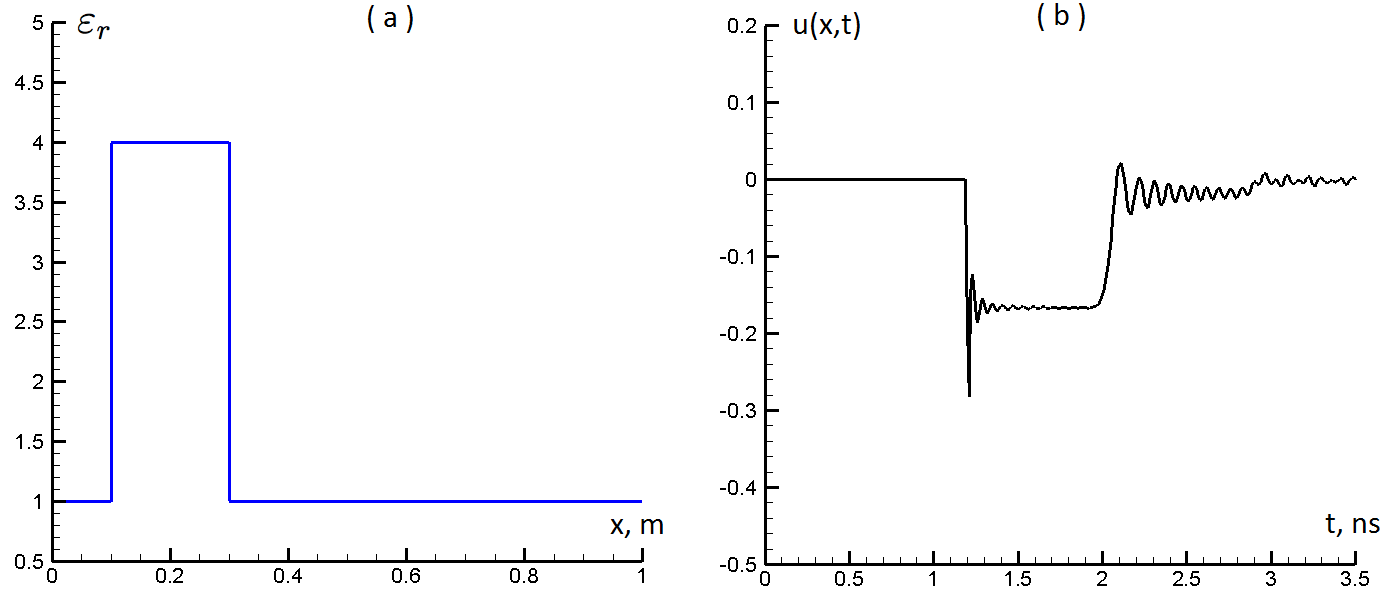}
\caption{\textsf{\ \emph{\ a) The graph of the function} $\protect\varepsilon%
_{r}(x)$ \emph{\ for which the data are computationally simulated via
solution of the forward problem (\protect\ref{2.3})-(\protect\ref{2.5}). b)
The computed function} $g_{1}(t):=g(t)-H(t-|x-x_{0}|)/2$ \emph{\ at the edge}
$\{x=0\}$, \emph{\ where the backscattering data are collected. The function}
$g(t)$ \emph{\ is defined in (\protect\ref{2.6})} }}
\label{fig_4}
\end{figure}

The function $g(t)$ models the backscattering data measured by the Forward
Looking Radar. We now solve the forward problem (\ref{2.3})-(\ref{2.5}) for
the case when the graph of the function $\varepsilon _{r}\left( x\right) $
is the one depicted on Fig. \ref{fig_4}-a). The resulting function $%
g_{1}(t):=g(t)-H\left( t-\left\vert x_{0}\right\vert \right) /2$ is depicted
on Fig. \ref{fig_4}-b). Here $H\left( t\right) $ is the Heaviside function
and $H\left( t-\left\vert x-x_{0}\right\vert \right) /2$ is the solution of
the problem (\ref{2.3})-(\ref{2.5}) for the case $\varepsilon _{r}\left(
x\right) \equiv 1.$

\begin{figure}[tb!]
\centering
\includegraphics[bb= 0 0 750 620, scale=0.3]{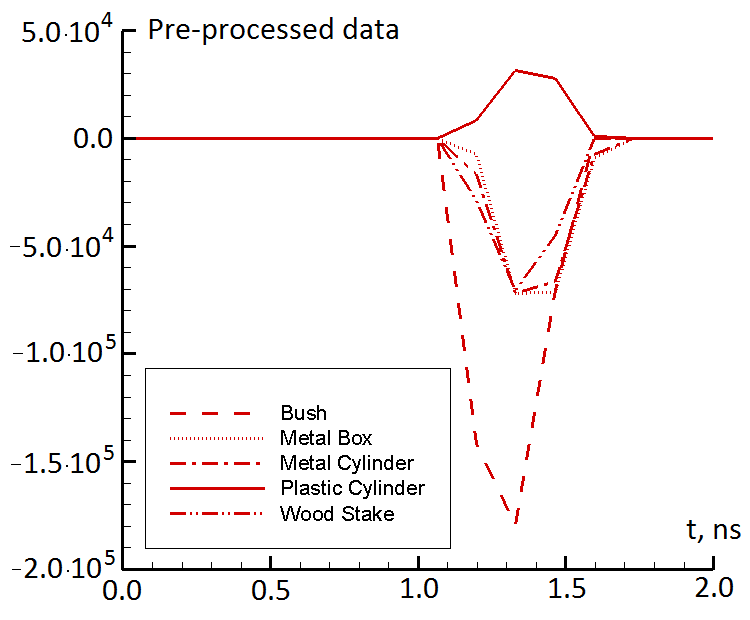}
\caption{\textsf{\ \emph{Pre-processed experimental data from the right
panel of Fig.~3. }}}
\label{fig_5}
\end{figure}

Comparison of Fig. \ \ref{fig_4}-b) with the data depicted on the right
panel of Fig. \ref{fig_3} confirms the above statement about a huge
discrepancy between experimental and computationally simulated data.
Therefore, we must change somehow the experimental data, i.e. pre-process
it. So that the pre-processes data would look somewhat similar with the
curve of Fig. \ref{fig_4}-b). Observe that this curve has basically only one
downward looking dent.\ If the step function of Fig. \ref{fig_4}-a) would be
downward looking, then we would have an upward looking step-like function on
the corresponding analog of Fig. \ref{fig_4}-b). Now, if a target is
standing in the air, then the background medium is air with $\varepsilon
_{r}(air)=1$ and $\varepsilon _{r}(target)>1.$\ Hence, it was decided in 
\cite{KBKSNF} to select only the first peak with the largest amplitude on
each experimental curve. The rest of the curve was set to zero. Still, in
order to take into account the information whether the target was in the air
or buried, a little bit different procedure of selecting peaks was
implemented. More precisely, the selected peak must be the earliest one with
the largest amplitude 
\begin{equation*}
\text{out of }\left\{ 
\begin{array}{c}
\text{all peaks for a buried target,} \\ 
\text{all downward looking peaks for a target in the air.}%
\end{array}%
\right.
\end{equation*}%
Also, since the time moment $\left\{ t=0\right\} $ was unknown to us, we set 
$t:=0$ to be such a point on the time axis, which is one (1) nanosecond to
the left from the beginning of the selected peak. Resulting superimposed
pre-processed curves are shown on Fig. \ref{fig_5}. It is clear now that the
condition $t\in \left( 0,\infty \right) $ in (\ref{2.6}) is not a
restrictive one, because each curve of Fig. \ref{fig_5} represents a
function with a compact support for $t\in \left( 0,\infty \right) .$ It is
also clear from comparison of Figures \ref{fig_4}-b) and \ref{fig_5} that
the pre-processed data are far from the range of the operator $A\left(
\varepsilon _{r}\right) :=u\left( 0,t\right) ,$ which should be inverted to
solve CIP1.

\subsubsection{Second stage: the choice of the calibration factor}

\label{sec:2.3.2}

The modulus of the curve of Fig. \ref{fig_4}-b) does not exceed 0.2. On the
other hand, amplitudes of peaks of experimentally measured curves of Fig. %
\ref{fig_3} are of the order of $10^{4}.$ This means that the data
correspond to $A\delta \left( x-x_{0}\right) $ in (\ref{2.4}), where the
number $A=const.>0$ is unknown. In our experience a similar situation takes
place for all experimental data, not necessarily for the current one. It is
clear therefore that in order to make the data suitable for the model (\ref%
{2.3})-(\ref{2.6}), they must be multiplied by a certain number, which we
call \textquotedblleft calibration factor" and denote $CF.$ A proper choice
of $CF$ is a non-trivial task. To have unbiased studies, it is \emph{%
important} that $CF$ must be the same for all targets.

In \cite{KBKSNF} the same calibration number $CF$ was chosen for all five
targets. It is important that because of the blind case, that choice was
unbiased. If we would have measured the dielectric constant of at least one
target in advance, then we would choose $CF$ in such a way that the number $%
R $ in (\ref{1}), which would be obtained via solving the above CIP, would
be close to the measured one. However, since dielectric constants were
unknown in advance, then another method of choice of $CF$ was used in \cite%
{KBKSNF}.\ Namely, consider the Laplace transform of the function $u\left(
x,t\right) ,$%
\begin{equation}
w(x,s)=\mathcal{L}\left( u\left( x,t\right) \right)
:=\int\limits_{0}^{\infty }u(x,t)e^{-st}dt,\ \ \ s\geq \underline{s}%
=const.>0.  \label{2.7}
\end{equation}%
We call the parameter $s$ pseudo frequency. Denote%
\begin{eqnarray}
\overline{g}\left( s\right) &=&\mathcal{L}g,  \label{600} \\
g_{0}\left( s\right) &=&\frac{1}{2}\mathcal{L}\left( H\left( t-\left\vert
x_{0}\right\vert \right) \right) =\frac{\exp \left( -s\left\vert
x_{0}\right\vert \right) }{2s},  \notag \\
\widehat{g}\left( s\right) &=&\overline{g}\left( s\right) -g_{0}\left(
s\right) .  \notag
\end{eqnarray}

Let $\widehat{g}_{sim}\left( s\right) $ be the Laplace transform of the
function depicted on Fig. \ref{fig_4}-b).\ Let $\widehat{g}_{bush}\left(
s\right) $ be the Laplace transform of the function, which corresponds to
bush on Fig. \ref{fig_5}. Then we multiply all experimental data by such a
calibration number $CF$ that values of functions $\widehat{g}_{sim}\left(
s\right) $ and $CF\cdot \widehat{g}_{bush}\left( s\right) $ would be not far
from each other for $s\in \left[ 1,5\right] $. It was shown in \cite{KBKSNF}
that $CF=10^{-7}$ was good for this goal. Furthermore, we have discovered
that the same value of $CF=10^{-7}$ provides similar behavior for Laplace
transforms of four other curves of Fig. \ref{fig_5}. In \cite{KBKSNF} we
have also varied the calibration factor $CF$ by 20\% as $CF_{1}=0.8\cdot
10^{-7},CF_{2}=1.2\cdot 10^{-7}$. In all three cases of $CF,CF_{1}$ and $%
CF_{2}$ the resulting values of dielectric constants were within tabulated
limits \cite{KBKSNF}. Hence, we have chosen in \cite{KBKSNF}%
\begin{equation}
CF=10^{-7}.  \label{602}
\end{equation}%
The choice of $CF$ for the GLK is a more delicate issue, and we discuss it
in section 5.1.

\section{Approximately Globally Convergent Method}

\label{sec:3}

In this section we briefly describe the 1-d version of AGCM for CIP1.
Details can be found in \cite{KBKSNF} as well as in section 6.9 of the book 
\cite{BK}. It was proved in \cite{KBKSNF} that the function $w(x,s)$ in (\ref%
{2.7}) is the solution of the following problem 
\begin{eqnarray}
w_{xx}-s^{2}\varepsilon _{r}\left( x\right) w &=&-\delta \left(
x-x_{0}\right) ,x\in \mathbb{R},\forall s\geq \underline{s},  \label{2.8} \\
\lim_{\left\vert x\right\vert \rightarrow \infty }w(x,s) &=&0.  \label{2.9}
\end{eqnarray}%
Denote%
\begin{equation}
w(0,s):=\overline{g}\left( s\right) ,\ \ s\geq \underline{s},  \label{2.10}
\end{equation}%
where the function $\overline{g}\left( s\right) $ is defined in (\ref{600}).
It was shown in \cite{KBKSNF} that 
\begin{equation}
w_{x}(0,s):=\rho (s)=s\overline{g}\left( s\right) -\exp (sx_{0}).
\label{2.11}
\end{equation}

\subsection{Integral differential equation}

\label{sec:3.1}

Consider the fundamental solution $w_{0}(x,s)$ of the problem (\ref{2.8}), (%
\ref{2.9}) for $\varepsilon _{r}(x)\equiv 1$. Then 
\begin{equation}
w_{0}(x,s)=\frac{1}{2s}\exp (-s|x-x_{0}|).  \label{2.12}
\end{equation}%
Since $w(x,s)>0$ \cite{KBKSNF}, we can consider $\ln \left[ w(x,s)\right] .$
Denote 
\begin{equation}
r(x,s)=\frac{\ln [(w/w_{0})(x,s)]}{s^{2}}.  \label{2.120}
\end{equation}%
Then (\ref{2.8})-(\ref{2.12}) imply that \cite{KBKSNF} 
\begin{eqnarray}
r_{xx}+s^{2}r_{x}^{2}-2sr_{x} &=&\varepsilon _{r}(x)-1,\ \ x>0,  \label{2.13}
\\[0.06in]
r(0,s) &=&\varphi _{0}(s),\ \ r_{x}(0,s)=\varphi _{1}(s),  \label{2.14}
\end{eqnarray}%
\begin{equation}
\varphi _{0}(s)=\frac{1}{s^{2}}\left[ \ln \overline{g}(s)-\ln (2s)\right] +%
\frac{x_{0}}{s},\ \ \varphi _{1}(s)=\frac{2}{s}-\frac{\exp (sx_{0})}{s^{2}%
\overline{g}(s)},  \label{2.15}
\end{equation}%
\begin{equation}
r_{x}(1,s)=0.  \label{2.16}
\end{equation}%
Differentiate equation (\ref{2.13}) with respect to $s$. Denote 
\begin{eqnarray}
q(x,s) &=&\partial _{s}r(x,s),\ \ \psi _{0}(s)=\varphi _{0}^{\prime }(s),\ \
\psi _{1}(s)=\varphi _{1}^{\prime }(s),  \label{2.17} \\
r(x,s) &=&-\int\limits_{s}^{\overline{s}}q(x,\tau )d\tau +V(x,\overline{s}),
\label{2.18} \\
V(x,\overline{s}) &=&\frac{1}{\overline{s}^{\,2}}\left\{ \ln \left[ w(x,%
\overline{s})\right] -\ln \left[ w_{0}(x,\overline{s})\right] \right\} =r(x,%
\overline{s}).  \label{2.19}
\end{eqnarray}%
Here $\overline{s}>0$ is a sufficiently large number which is chosen in
numerical experiments. We call $V(x,\overline{s})$ the \emph{tail function}.
Actually $\overline{s}$ is one of regularization parameters of this method.
We have 
\begin{eqnarray}
V(x,\overline{s}) &=&\frac{p(x)}{\overline{s}}+O\left( \frac{1}{\overline{s}%
^{2}}\right) ,\text{ }\overline{s}\rightarrow \infty ,  \label{2.20} \\
q(x,\overline{s}) &=&-\frac{p(x)}{\overline{s}^{2}}+O\left( \frac{1}{%
\overline{s}^{3}}\right) ,\ \overline{s}\rightarrow \infty .  \label{2.201}
\end{eqnarray}%
Using (\ref{2.13})-(\ref{2.19}), we obtain

\begin{equation*}
q_{xx}-2s^{2}q_{x}\int\limits_{s}^{\overline{s}}q_{x}(x,\tau )d\tau +2s\left[
\int\limits_{s}^{\overline{s}}q_{x}(x,\tau )d\tau \right] ^{2}-2sq_{x}+2\int%
\limits_{s}^{\overline{s}}q_{x}(x,\tau )\,d\tau
\end{equation*}%
\begin{equation}
+2s^{2}q_{x}V_{x}-4sV_{x}\int\limits_{s}^{\overline{s}}q_{x}(x,\tau )d\tau
+2sV_{x}^{2}-2V_{x}=0,\ \ s\in \lbrack \underline{s},\overline{s}],
\label{2.22}
\end{equation}

\begin{equation}
q(0,s)=\psi _{0}(s),\ \ q_{x}(0,s)=\psi _{1}(s),\ \ q_{x}(1,s)=0,\ \ s\in
\lbrack \underline{s},\overline{s}].  \label{2.23}
\end{equation}

The main difficulty of this numerical method is in the solution of the
problem (\ref{2.22}), (\ref{2.23}). Equation (\ref{2.22}) has two unknown
functions $q$ and $V$. To approximate both of them, we use a scheme, which
is based on both inner and outer iterations \cite{KuzhKl,KBKSNF}. First,
given an approximation for $V$, we update $q$ via solution of a boundary
value problem on the interval $x\in \left( 0,1\right) $. This is our inner
iteration. Next, we update the unknown coefficient $\varepsilon _{r}(x)$ and
solve the forward problem (\ref{2.8}), (\ref{2.9}) for $s:=\overline{s}$ on
the entire real line $\mathbb{R}$ with this updated coefficient $\varepsilon
_{r}(x)$. Next, we update the tail function $V\left( x,\overline{s}\right) $
via (\ref{2.19}). If both functions $q$ and $V$ are approximated well by
functions $\widetilde{q}$ and $\widetilde{V}$ respectively via this
numerical procedure, then we use backwards calculations (\ref{2.18}), (\ref%
{2.13}) to approximate the unknown coefficient $\widetilde{\varepsilon }%
_{r}(x),$%
\begin{eqnarray}
\widetilde{r}(x,s) &=&-\int\limits_{s}^{\overline{s}}\widetilde{q}(x,\tau
)d\tau +\widetilde{V}(x,\overline{s}),  \label{2.24} \\
\widetilde{\varepsilon }_{r}\left( x\right) &=&\left\{ 
\begin{array}{c}
1+\widetilde{r}_{xx}+s^{2}\widetilde{r}_{x}^{2}-2s\widetilde{r}_{x},\ \ x\in
(0,1), \\[4pt] 
1,\ \ \ x\notin \left( 0,1\right) .%
\end{array}%
\right.  \label{2.25}
\end{eqnarray}

\subsection{Numerical procedure}

\label{sec:3.2}

Consider a partition of the interval $\left[ \underline{s},\overline{s}%
\right] $ into $N$ small subintervals with the grid step size $h>0$ and
assume that the function $q\left( x,s\right) $ is piecewise constant with
respect to $s$, 
\begin{equation*}
\underline{s}=s_{N}<s_{N-1}<...<s_{0}=\overline{s},s_{i-1}-s_{i}=h;\text{ }%
q\left( x,s\right) =q_{n}\left( x\right) ,\text{ for }s\in \left(
s_{n},s_{n-1}\right] .
\end{equation*}%
For each subinterval $\left( s_{n},s_{n-1}\right] $ we obtain a differential
equation for the function $q_{n}\left( x\right) .$ We assign for convenience
of notations $q_{0}:\equiv 0.$ For each $n$ we iterate to improve tails.
Next, we go to the next $n$. This way we obtain functions $q_{n,k},V_{n,k}.$
The equation for the pair $\left( q_{n,k},V_{n,k}\right) $ is

\begin{equation*}
q_{n,k}^{\prime \prime }-\left( A_{1,n}h\sum\limits_{j=0}^{n-1}q_{j}^{\prime
}-A_{1,n}V_{n,k}^{\prime }-2A_{2,n}\right) q_{n,k}^{\prime }=
\end{equation*}%
\begin{equation}
-A_{2,n}h^{2}\left( \sum\limits_{j=0}^{n-1}q_{j}^{\prime }\right)
^{2}+2h\sum\limits_{j=0}^{n-1}q_{j}^{\prime }+2A_{2,n}V_{n,k}^{\prime
}\left( h\sum\limits_{j=0}^{n-1}q_{j}^{\prime }\right)  \label{2.27}
\end{equation}%
\begin{equation*}
-A_{2,n}\left( V_{n,k}^{\prime }\right) ^{2}+2A_{2,n}V_{n,k}^{\prime },\text{
}x\in \left( 0,1\right) ,n\in \left[ 1,N\right] ,k\in \left[ 1,m\right] ,
\end{equation*}%
where the number $m$ should be chosen in numerical experiments. The boundary
conditions are generated by (\ref{2.23}),%
\begin{equation}
q_{n,k}\left( 0\right) =\psi _{0,n},q_{n,k}^{\prime }\left( 0\right) =\psi
_{1,n},q_{n,k}^{\prime }\left( 1\right) =0.  \label{2.28}
\end{equation}%
Boundary conditions (\ref{2.28}) are over-determined ones. On the other
hand, our attempts to use only two out of three boundary conditions (\ref%
{2.28}) never led to acceptable results \cite{KuzhKl}. Thus, given the
function $V_{n,k}^{\prime }\left( x\right) ,$ we use all three conditions (%
\ref{2.28}) and find an approximate solution $q_{n,k}\left( x\right) $ of
the problem (\ref{2.27}), (\ref{2.28}) via the Quasi-Reversibility Method
(QRM), see subsection 3.3. We refer to the book \cite{LL} for the
originating work about the QRM and to \cite{BK,Bourg,Kl,KuzhKl,KBKSNF} for
some follow up publications. QRM is well suited for solving over-determined
boundary value problems. Convergence of QRM is proved via Carleman estimates 
\cite{Kl}.

The choice of the first tail function $V_{1,1}\left( x\right) $ is described
in subsection 3.3. Let $n\geq 1.$ Suppose that for $j=0,...n-1$ functions $%
q_{j}\left( x\right) ,V_{j}\left( x\right) $ are constructed. We now
construct functions $q_{n,k},V_{n,k}$ for $k=1,...,m.$ We set $V_{n,1}\left(
x\right) :=V_{n-1}\left( x\right) .$ If the function $V_{n,k}\left( x\right) 
$ is given, then we find an approximate solution $q_{n,k}\left( x\right) $
of the problem (\ref{2.27}), (\ref{2.28}). Next, we find the approximation $%
\varepsilon _{r}^{\left( n,k\right) }$ for the unknown coefficient $%
\varepsilon _{r}\left( x\right) $ via discrete analogs of (\ref{2.24}), (\ref%
{2.25}), 
\begin{equation*}
r_{n,k}\left( x\right) =-hq_{n,k}-h\sum\limits_{j=0}^{n-1}q_{j}+V_{n,k},%
\text{ }x\in \left( 0,1\right) ,
\end{equation*}%
\begin{equation*}
\varepsilon _{r}^{\left( n,k\right) }\left( x\right) =\left\{ 
\begin{array}{c}
1+r_{n,k}^{\prime \prime }\left( x\right) +s_{n}^{2}\left[ r_{n,k}^{\prime
}\left( x\right) \right] ^{2}-2s_{n}r_{n,k}^{\prime }\left( x\right) ,x\in
\left( 0,1\right) , \\ 
1,x\notin \left( 0,1\right) .%
\end{array}%
\right. \text{ }
\end{equation*}%
Next, we solve the forward problem (\ref{2.8}), (\ref{2.9}) with $%
\varepsilon _{r}\left( x\right) :=\varepsilon _{r}^{\left( n,k\right)
}\left( x\right) ,s:=\overline{s}$ and find the function $w_{n,k+1}\left( x,%
\overline{s}\right) $ this way. After this we update the tail via formula (%
\ref{2.19}) in which $w\left( x,\overline{s}\right) :=w_{n,k+1}\left( x,%
\overline{s}\right) .$ This way we obtain a new tail $V_{n,k+1}\left(
x\right) .$ Similarly we continue iterating with respect to tails $m$ times.
Next, we set 
\begin{equation*}
q_{n}\left( x\right) :=q_{n,m}\left( x\right) ,\text{ }V_{n}\left( x\right)
:=V_{n+1,1}\left( x\right) :=V_{n,m}\left( x\right) ,\text{ }\varepsilon
_{r}^{\left( n\right) }\left( x\right) :=\varepsilon _{r}^{\left( n,m\right)
}\left( x\right) ,
\end{equation*}%
replace $n$ with $n+1$ and repeat this process. By \cite{KuzhKl} we continue
this process until 
\begin{equation}
\text{either }\left\Vert \varepsilon _{r}^{\left( n\right) }-\varepsilon
_{r}^{\left( n-1\right) }\right\Vert _{L_{2}\left( 0,1\right) }\leq 10^{-5}%
\text{ or }\left\Vert \nabla J_{\alpha }\left( q_{n,k}\right) \right\Vert
_{L_{2}\left( 0,1\right) }\geq 10^{5},  \label{2.29}
\end{equation}%
where the functional $J_{\alpha }\left( q_{n,k}\right) $ is defined in (\ref%
{2.30}) of subsection 3.3. Here the norm in the space $L_{2}\left(
0,1\right) $ is understood in the discrete sense. In the case when the
second inequality (\ref{2.29}) is satisfied, we stop at the previous
iteration taking $\varepsilon _{r}^{\left( n,k-1\right) }\left( x\right) $
as our solution. If neither of two conditions (\ref{2.24}) is not reached at 
$n:=N,$ then we repeat the above sweep over the interval $\left[ \underline{s%
},\overline{s}\right] $ taking the tail $V_{N}\left( x\right) $ as the new
tail $V_{1,1}\left( x\right) .$ Usually at least one of conditions (\ref%
{2.29}) is reached at most on the third or on the fourth sweep, and the
process stops then.

\subsection{Computing functions $q_{n,k}(x)$, $V_{1,1}(x)$}

\label{sec:3.3}

As it was mentioned in Subsection 3.2, we compute approximations for
functions $q_{n,k}\left( x\right) $ via QRM. Let $L\left( q_{n,k}\right)
\left( x\right) $ be the operator in the left hand side of equation (\ref%
{2.27}) and $P_{n,k}\left( x\right) $ its right hand side. Let $\alpha \in
\left( 0,1\right) $ be the regularization parameter. The QRM minimizes the
following Tikhonov regularization functional%
\begin{equation}
J_{\alpha }\left( q_{n,k}\right) =\left\Vert L_{n,k}\left( q_{n,k}\right)
-P_{n,k}\right\Vert _{L_{2}\left( 0,1\right) }^{2}+\alpha \left\Vert
q_{n,k}\right\Vert _{H^{2}\left( 0,1\right) }^{2},  \label{2.30}
\end{equation}%
subject to boundary conditions (\ref{2.28}).

We now describe an important step of choosing the first tail function $%
V_{1,1}\left( x\right) .$ Consider the asymptotic behavior (\ref{2.20}) and (%
\ref{2.201}) of functions $V\left( x,\overline{s}\right) $ and $q\left( x,%
\overline{s}\right) $ with respect to the truncation pseudo frequency $%
\overline{s}\rightarrow \infty .$ We truncate in (\ref{2.20}) and (\ref%
{2.201}) terms $O\left( 1/\overline{s}^{2}\right) $ and $O\left( 1/\overline{%
s}^{3}\right) $ respectively. This is somewhat similar with the definition
of geometrical optics as a high frequency approximation of the solution of
the Helmholtz equation. Hence, for sufficiently large $\overline{s}$ 
\begin{equation}
V\left( x,\overline{s}\right) \approx \frac{p\left( x\right) }{\overline{s}}%
,\ \ \ q\left( x,\overline{s}\right) \approx -\frac{p\left( x\right) }{%
\overline{s}^{2}}.  \label{2.31}
\end{equation}%
Hence, setting in equation (\ref{2.22}) $s:=\overline{s}$ and using (\ref%
{2.31}), we obtain the following \emph{approximate} equation for the
function $p\left( x\right) $%
\begin{equation}
p^{\prime \prime }\left( x\right) =0,\ \ \ \ x\in \left( 0,1\right) .
\label{2.32}
\end{equation}%
Boundary conditions for the function $p\left( x\right) $ can be easily
derived from (\ref{2.23}) and (\ref{2.31}) as 
\begin{equation}
p\left( 0\right) =-\overline{s}^{2}\psi \left( \overline{s}\right) ,\text{ }%
p^{\prime }\left( 0\right) =-\overline{s}^{2}\psi _{1}\left( \overline{s}%
\right) ,\text{ }p^{\prime }\left( 1\right) =0.  \label{2.33}
\end{equation}%
We find an approximate solution $\overline{p}\left( x\right) $ of the
problem (\ref{2.32}), (\ref{2.33}) via the QRM. Next, we set for the first
tail function%
\begin{equation}
V_{1,1}\left( x\right) :=\frac{\overline{p}\left( x\right) }{\overline{s}}.
\label{2.34}
\end{equation}

Theorem 1 is a simplified version of Theorem 6.1 of \cite{KBKSNF}, also see
Theorem 6.7 of \cite{BK} for the 3-d case.

\textbf{Theorem 1}. \emph{Let the function }$\varepsilon _{r}^{\ast }\left(
x\right) $\emph{\ satisfying conditions (\ref{2.1}), (\ref{2.2}) be the
exact solution of our CIP for the noiseless data }$g^{\ast }\left( t\right) $%
\emph{\ in (\ref{2.6}). Fix the truncation pseudo frequency }$\overline{s}%
>1. $\emph{\ Let the first tail function }$V_{1,1}\left( x\right) $\emph{\
be defined via (\ref{2.32})-(\ref{2.34}). Let }$\gamma \in \left( 0,1\right) 
$\emph{\ be the level of the error in the boundary data, i.e.}%
\begin{equation*}
\left\vert \psi _{0}\left( s\right) -\psi _{0}^{\ast }\left( s\right)
\right\vert \leq \gamma ,\text{ }\left\vert \psi _{1}\left( s\right) -\psi
_{1}^{\ast }\left( s\right) \right\vert \leq \gamma ,\text{for }s\in \left[ 
\underline{s},\overline{s}\right] ,
\end{equation*}%
\emph{where functions }$\psi _{0}\left( s\right) ,\psi _{1}\left( s\right) $%
\emph{\ depend on the function }$g\left( t\right) $\emph{\ in (\ref{2.6})
via (\ref{600}), (\ref{2.15}), (\ref{2.17}) and functions }$\psi _{0}^{\ast
}\left( s\right) ,\psi _{1}^{\ast }\left( s\right) $\emph{\ depend on the
noiseless data }$g^{\ast }\left( t\right) $\emph{\ in the same way. Let }$%
\sqrt{\alpha }=\gamma $ \emph{and }$\widetilde{h}=\max \left( \gamma
,h\right) $\emph{. Let }$Q$\emph{\ be the total number of functions }$%
\varepsilon _{r}^{\left( n,k\right) }$\emph{\ computed in the above
algorithm. Then there exists a constant }$D=D\left( x_{0},d_{0},d_{1},%
\overline{s}\right) >1$\emph{\ such that if the number }$\widetilde{h}$\emph{%
\ is so small that }%
\begin{equation}
\widetilde{h}<\frac{1}{D^{Q}}\emph{,}  \label{2.35}
\end{equation}%
\emph{\ then the following convergence estimate is valid }%
\begin{equation}
\left\Vert \varepsilon _{r}^{\left( n,k\right) }-\varepsilon _{r}^{\ast
}\right\Vert _{L_{2}\left( 0,1\right) }\leq \widetilde{h}^{\omega },
\label{2.36}
\end{equation}%
\emph{where the number }$\omega \in (0,1)$\emph{\ is independent on} $n,k,%
\widetilde{h},\varepsilon _{r}^{(n,k)},\varepsilon _{r}^{\ast }.$

Theorem 1 guarantees that if the total number $Q$ of computed functions $%
\varepsilon _{r}^{\left( n,k\right) }$ is fixed and error parameters $\sigma
,h$ are sufficiently small, then iterative solutions $\varepsilon
_{r}^{\left( n,k\right) }\left( x\right) $ are sufficiently close to the
exact solution $\varepsilon _{r}^{\ast },$ and this closeness is defined by
the error parameters. Estimate (\ref{2.36}) guarantees the stability of AGCM
with respect to a small error in the data. Therefore the total number of
iterations $Q$ can be considered as one of regularization parameters of our
process.\ Two other regularization parameters are the numbers $\overline{s}$
and $\alpha $. The combination of inequalities (\ref{2.35}) and (\ref{2.36})
has a direct analog in the inequality of Lemma 6.2 on page 156 of the book 
\cite{EHN} for classical Landweber iterations, which are defined for a
substantially different ill-posed problem. Indeed, it is stated on page 157
of the book \cite{EHN} that the number of iterations can serve as a
regularization parameter for an ill-posed problem.

The \emph{main advantage }of Theorem 1 is that even though it does not
require any advanced knowledge of a small neighborhood of the exact solution 
$\varepsilon _{r}^{\ast },$ it still guarantees the most important property
of this algorithm: the property of delivering at least one point in this
neighborhood. In principle, (\ref{2.36}) guarantees that one can take the
function $\varepsilon _{r}^{\left( 1,1\right) }$ resulting from just the
first tail $V_{1,1}\left( x\right) $ as an approximate solution. In other
words, already on the first iteration we obtain a good approximation. Next,
one can proceed either of two ways.\ First, one can continue iterations as
long as the total number $Q$ of iterations satisfies inequality (\ref{2.35}%
). In doing so, one can choose an optimal number of iterations, based on
numerical experience, see (\ref{2.29}) for our stopping rule. Second, one
can choose either of functions $\varepsilon _{r}^{\left( n,k\right) }$ with $%
nk\leq Q$ as an approximate solution and then refine it using one of locally
convergent numerical methods. This is a two-stage numerical procedure, which
was developed in \cite{BK2,BK} and some other publications about AGCM. An
effective method of this sort is, e.g. Adaptive Finite Element Method
(adaptivity), which minimizes the Tikhonov functional on a sequence of
locally refined finite element meshes, see, e.g. \cite{b,BK4} and Chapter 4
of\ the book \cite{BK}.

In the numerical implementation of AGCM the functional $J_{\alpha }\left(
q_{n,k}\right) $ in (\ref{2.30}) was written via finite differences and was
minimized with respect to the values of the function $q_{n,k}$ at grid
points. Conjugate gradient method was used for the minimization. Convergence
of this method to the unique minimizer of $J_{\alpha }\left( q_{n,k}\right) $
follows from the strong convexity of $J_{\alpha }\left( q_{n,k}\right) ,$
which, in turn follows from Lemma 5.2 of \cite{KuzhKl}. The $s-$interval was 
$s\in \left[ \underline{s},\overline{s}\right] =\left[ 0.5,12\right] .$ The
length of each small subinterval $\left( s_{n},s_{n-1}\right) $ was $h=0.5$.
We took $\alpha =0.04$ in (\ref{2.30}). For each $n=1,...,N$ we have
calculated functions $q_{n,k}$ for $k=1,...,10:=m.$ We refer to \cite%
{KuzhKl,KBKSNF} for details of the numerical implementation of AGCM for this
case.

\subsection{Numerical results for AGCM}

\label{sec:3.5}

\begin{figure}[tb!]
\centering
\includegraphics[bb= 0 0 2050 630, scale=0.22]{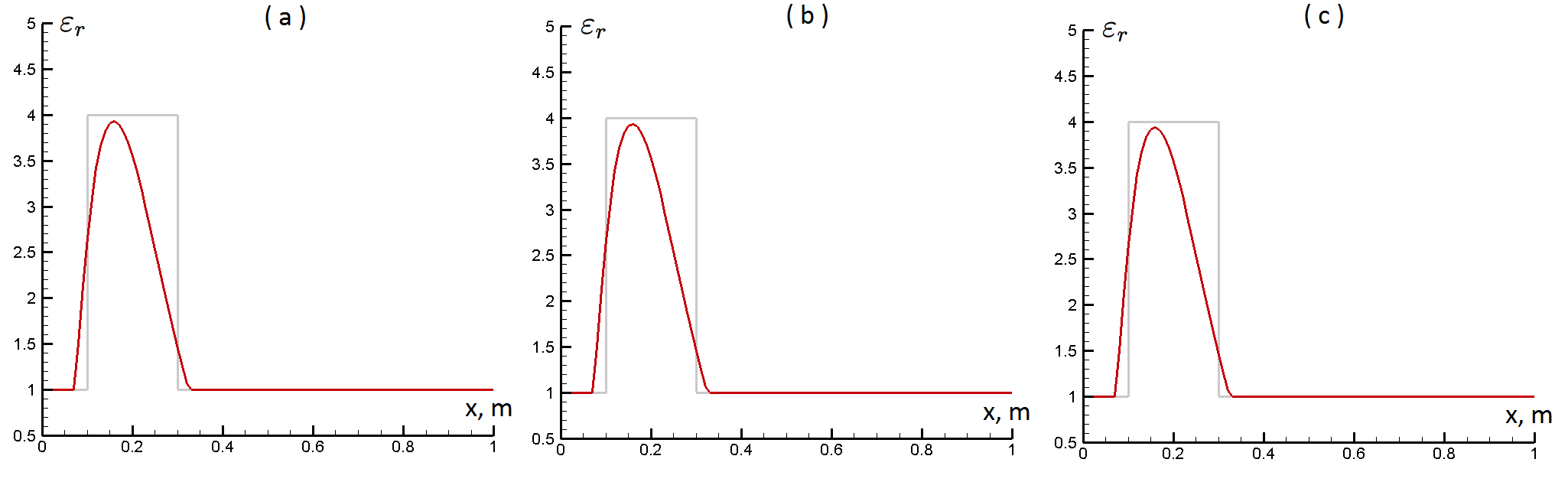}
\caption{\textsf{\ \emph{Images from computationally simulated data for
different values of the random noise in the data, see (\protect\ref{2.37}).
The true target is the same as on Fig.~\protect\ref{fig_4}-a). a) 0\% noise,
b) 5\% noise, c) 10\% noise.} }}
\label{fig_6}
\end{figure}

\textbf{Test 1}. We now test AGCM for computationally simulated data for the
case of the function $\varepsilon _{r}\left( x\right) $ displayed on Fig. %
\ref{fig_4}-a). We have introduced the multiplicative random noise in the
function $g\left( t\right) $ in (\ref{2.6}) as%
\begin{equation}
g_{\xi }\left( t_{i}\right) =g\left( t_{i}\right) \left( 1+\sigma \xi
_{i}\right) ,  \label{2.37}
\end{equation}%
where $t=t_{i}$ is the discrete value number $i$ of the variable $t\in
\left( 0,T\right) $, which was used in data simulations, $\sigma \in \left(
0,1\right) $ is the noise level and $\xi _{i}$ is the value number $i$ of
the random variable $\xi \in \left( -1,1\right) $. Hence, e.g. $\sigma =0.05$
means 5\% of the noise level in the data. Because of Fig. \ref{fig_4}-b) as
well as because of the rapid decay of the kernel $e^{-st}$ of the Laplace
transform (\ref{2.7}), we took $T=4.$ Images with 0\% , 5\% and 10\% noise
in (\ref{2.37}) are displayed on Fig. \ref{fig_6}. One can observe that both
the location of the abnormality and the maximal value of the coefficient $%
\varepsilon _{r}\left( x\right) $ are imaged with a good accuracy for all
three cases.

\begin{figure}[tb!]
\centering
\includegraphics[bb= 0 0 660 610, scale=0.22]{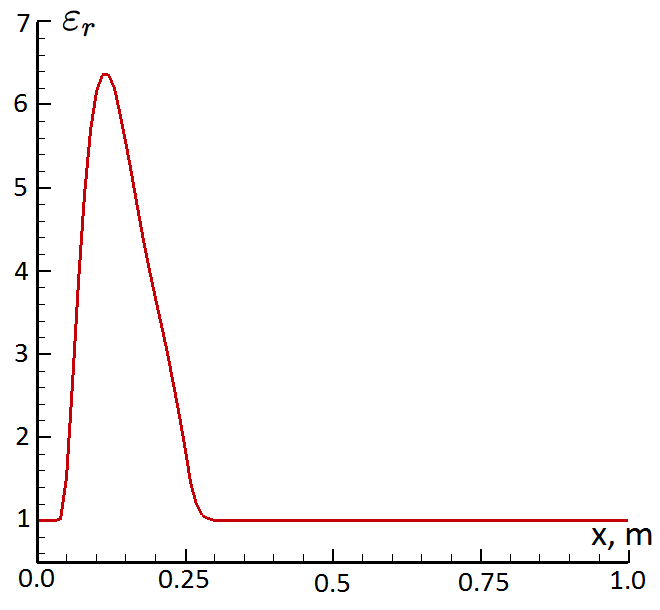}
\caption{\textsf{\ \emph{The blindly computed function} $R(x) $\emph{\ for
bush, see the fourth row of Fig.~\protect\ref{fig_3}. The computed value} $%
\max R(x)=\overline{R}=6.4=\protect\varepsilon_{r}(\text{bush}) $\emph{\ is
within tabulated limits of } $[3,20]$ \emph{\protect\cite{vegetation}} }}
\label{fig_7}
\end{figure}

\textbf{Test 2}. We now show the image using experimental data for case of
\textquotedblleft bush". This case is the most difficult one, since bush is
highly heterogeneous. We remind that because our data are severely
under-determined ones, the calculated function $R\left( x\right) $ in (\ref%
{1}) is a sort of a weighted average value of the ratio (\ref{1}) over the
volume of a target. Next, we have applied formulas (\ref{2}), (\ref{3}) to
find $\varepsilon _{r}\left( target\right) .$ We do not reproduce images of
other four available cases of real data, since they are published in \cite%
{KBKSNF}. Fig. \ref{fig_7} displays this image. Hence, $\overline{R}=6.4$ in
this case. As one can see from Fig. \ref{fig_3}, bush was standing in the
air. Hence, in this case $\varepsilon _{r}\left( bckgr\right) =1.$ We
conclude therefore that the computed value $\varepsilon _{r}\left( \text{bush%
}\right) =6.4.$ On the other hand it is clear from\ \cite{vegetation} that
the tabulated value is $\varepsilon _{r}\left( \text{bush}\right) \in \left[
3,20\right] .$ Therefore, our blindly computed value of the dielectric
constant of bush is within tabulated limits.

Table 1 \cite{KBKSNF} shows computed values for all five cases of
experimental data. The background value of the dielectric constant of the
dry sand, in which three out of five targets were buried, was between 3 and
5 \cite{Table1}. It is clear from Fig. \ref{fig_3} which targets were buried
and which ones were standing in the air. The dielectric constant for metals
is not defined in Physics. Nevertheless, it was established in numerical
experiments in \cite{KBKSNF} that as soon as the value of the dielectric
constant of a target $\varepsilon _{r}\geq 10,$ the backscattering electric
wave from this target is about the same as the one from a metallic target.
Therefore, the so-called \emph{appearing dielectric constant }for metals was
introduced in \cite{KBKSNF}, i.e. this is such a dielectric constant, which
is between 10 and 30. Thus we assign for this constant%
\begin{equation}
\varepsilon _{r}\left( \text{metals}\right) \in \left[ 10,30\right] .
\label{2.38}
\end{equation}%
Tabulated values of dielectric constants of the wood stake and plastic
cylinder in Table 1 are taken from Tables \cite{Table1,Table2}. As to the
bush, see \cite{vegetation}. The tabulated value of $\varepsilon _{r}\left( 
\text{metal}\right) $ is taken from (\ref{2.38}). Note that published values
of dielectric constants are given within certain intervals. Therefore, by (%
\ref{3}),\emph{\ }values of computed and tabulated $\varepsilon _{r}\left( 
\text{target}\right) $ in Table 1 are also given within certain intervals
denoted by brackets. This is unlike computed values of $\overline{R}.$

\begin{center}
\begin{table}[h!]
\caption{Blindly computed dielectric constants for five targets of Fig.~%
\protect\ref{fig_3} as in (\protect\ref{602}), by the globally convergent
numerical method versus tabulated ones. ``1'' in the second column means
that the target was in the air and ``$[3,5]$'' means that it was buried in
the dry sand whose dielectric constant is between 3 and 5 \protect\cite%
{Table1}. }%
\begin{tabular}{|l|l|l|l|l|}
\hline
target & $\overline{R}$ & $\varepsilon _{r}(bckgr)$ & computed $\varepsilon
_{r}\left( \text{target}\right) $ & tabulated $\varepsilon _{r}\left( \text{%
target}\right) $ \\ \hline
metal cylinder & $4.3$ & $[3,5] $ & $[12.9,21.4] $ & $[10,30]$, see (\ref%
{2.38}) \\ \hline
metal box & $3.8$ & $[3,5]$ & $[11.4,19] $ & $[10,30]$, see (\ref{2.38}) \\ 
\hline
wood stake & $3.8$ & $1$ & $3.8$ & $[2,6]$, see \cite{Table1} \\ \hline
bush (clutter) & $6.4$ & $1$ & $6.4$ & $[3,20]$, see \cite{vegetation} \\ 
\hline
plastic cylinder & $0.28$ & $[3,5] $ & $[0.84,1.4]$ & $1.2,$ see \cite%
{Table2} \\ \hline
\end{tabular}%
\end{table}
\end{center}

\textbf{Conclusion}. Since dielectric constants of targets were not measured
in experiments, the only thing we can do is to compare computed values of $%
\varepsilon _{r}\left( \text{target}\right) $ with tabulated ones. The \emph{%
most important} conclusion, which can be drawn from Table~1, is that in five
out of five blind cases computed values of dielectric constants of targets
were well within tabulated limits. We remind that the variation of the
calibration factor $CF=10^{-7}$ in (\ref{602}) by 20\% has also resulted in
values being well within tabulated limits.

\section{The Gelfand-Levitan-Krein Integral Equation (GLK)}

\label{sec:4}

\subsection{Forward and inverse problems}

\label{sec:4.1}

The forward problem in this case is 
\begin{eqnarray}
\varepsilon _{r}(x)u_{tt}=u_{xx}, &\ &x>0,\ \ t>0,  \label{4.1} \\[0.08in]
u\left( x,0\right) =u_{t}\left( x,0\right) =0, &&  \label{4.2} \\[0.08in]
u_{x}|_{x=0}=\delta (t). &&  \label{4.3}
\end{eqnarray}%
\textbf{Coefficient Inverse Problem 2 (CIP2).} Determine the function $%
\varepsilon _{r}(x)$ in (\ref{4.1}), assuming that the following function $%
f(t)$ is known 
\begin{equation}
u(0,t)=f(t),\ \ t\geq 0.  \label{4.4}
\end{equation}

Without loss of generality, we assume that 
\begin{equation}
\varepsilon _{r}(0)=1.  \label{4.5}
\end{equation}%
Change the spatial variable as 
\begin{equation}
x\Leftrightarrow z=\tau (x)=\int\limits_{0}^{x}\sqrt{\varepsilon _{r}\left(
\tau \right) }d\tau .  \label{4.6}
\end{equation}%
Hence, taking into account (\ref{4.5}), we obtain that (\ref{4.4}) holds at $%
z=0$ and (\ref{4.1})-(\ref{4.3}) imply that 
\begin{eqnarray}
u_{tt}=u_{zz}-\frac{\sigma ^{\prime }(z)}{\sigma (z)}u_{z}, &\ &z>0,\ \ t>0,
\label{4.7} \\[0.03in]
u\left( z,0\right) =u_{z}\left( z,0\right) =0, &&  \label{4.8}
\end{eqnarray}%
\begin{eqnarray}
u_{z}|_{z=0} &=&\delta (t),  \label{4.9} \\
\sigma \left( z\right) &=&\frac{1}{\sqrt{\varepsilon _{r}(\tau ^{-1}(z))}}.
\label{4.10}
\end{eqnarray}%
Consider even extensions of functions $\sigma (z)$ and $u(z,t)$ in $\{z<0\}$%
. To do this, we set without changing notations $\sigma (z)=\sigma
(-z),u(z,t)=u(-z,t)$ for $z<0.$ It follows from \cite{Blag} that we obtain
then the following Cauchy problem 
\begin{eqnarray*}
u_{tt} &=&u_{zz}-\frac{\sigma ^{\prime }(z)}{\sigma (z)}u_{z},\ \ \ z\in 
\mathbb{R},\ \ t>0, \\[2pt]
u(z,0) &=&0,\ \ u_{t}(z,0)=-2\delta (z).
\end{eqnarray*}%
Hence, 
\begin{equation}
u(z,t)=-\sqrt{\sigma (z)}\cdot H(t-\left\vert z\right\vert )+\tilde{u}(z,t),
\label{4.13}
\end{equation}%
where the function $\tilde{u}(z,t)$ is continuous for $\left( z,t\right) \in 
\mathbb{R}\times \left( 0,T\right) ,\forall T>0$ and $\tilde{u}(z,t)=0$ for $%
t\leq \left\vert z\right\vert .$ Hence, (\ref{4.10}) and (\ref{4.13}) imply
that $f(0^{+})\!=\!u(0^{+},0^{+})\!=\!-\sqrt{\sigma \left( 0\right) }.$
Combining this with (\ref{4.5}), we obtain 
\begin{equation}
f(0^{+})\!=\!-1.  \label{4.14}
\end{equation}%
Let 
\begin{equation*}
\hat{f}(t)=\left\{ 
\begin{array}{rl}
f(t), & t>0 \\[2mm] 
-f(-t) & t<0%
\end{array}%
.\right.
\end{equation*}%
Hence, the function $\hat{f}(t)$ is odd and its first derivative $\hat{f}%
^{\prime }(t)$ is even. Choose a number $T>0.$ Taking into account (\ref%
{4.14}) and using results of \cite{Blag,KabSSat}, we obtain that the GLK
equation has the form%
\begin{equation}
w(z,t)-\frac{1}{2}\int\limits_{-z}^{z}\hat{f}^{\prime }(t-\tau )w(z,\tau
)d\tau =\frac{1}{2},\ \ \ t\in \lbrack -z,z],\ \ \forall z\in \lbrack 0,T/2].
\label{4.15}
\end{equation}%
This is not a Volterra equation, because $z\in \lbrack 0,T/2]$ is a
parameter here. Hence, equation (\ref{4.15}) should be solved for all values
of the parameter $z\in \lbrack 0,T/2].$ The solution of this equation is
connected with the function $\sigma (z)$ via \cite{Blag,KabSSat}%
\begin{equation}
\lim_{t\rightarrow z^{-}}w\left( z,t\right) :=w(z,z^{-})=\frac{1}{2\sqrt{%
\sigma (z)}},\ \ z\in \lbrack 0,T/2].  \label{4.16}
\end{equation}%
Hence, by (\ref{4.10}) and (\ref{4.16}) 
\begin{equation}
\varepsilon _{r}\left( \tau ^{-1}\left( z\right) \right) =\frac{1}{%
16w^{4}(z,z^{-})}.  \label{4.17}
\end{equation}%
Therefore, in order to solve CIP2, one needs first to solve equation (\ref%
{4.15}) for all $z\in \lbrack 0,T/2].$ Next, one needs to reconstruct the
function $\varepsilon _{r}\left( \tau ^{-1}\left( z\right) \right) $ via (%
\ref{4.17}). Finally, one needs to change variables backwards transforming $%
z $ into $x$ via inverting (\ref{4.6}).

\textbf{Theorem 2 } \cite{Blag,Rom}.\emph{\ Let the function }$f(t)$\emph{\
be given for }$t\in \lbrack 0,T]$\emph{\ for a certain value of }$T>0$\emph{%
. Suppose that there exists unique function }$\varepsilon _{r}(x)$\emph{\ of
CIP2 satisfying conditions (\ref{2.1}), (\ref{2.2}), in which }$\mathbb{R}$%
\emph{\ is replaced with }$\left\{ x\geq 0\right\} .$\emph{\ Also, assume
that condition (\ref{4.5}) holds. Let }$\tau (z)$\emph{\ be the function
defined in (\ref{4.6})}.\emph{\ Then solution of the GLK integral equation (%
\ref{4.15}) exists for each }$z\in \lbrack 0,T/2]$\emph{\ and is unique. And
vice versa: if the function }$f(t)$\emph{\ is known for all }$t\in \lbrack
0,T]$\emph{\ and the integral equation (\ref{4.15}) has unique solution for
each }$z\in \lbrack 0,T/2]$\emph{, then CIP2 has the unique solution }$%
\varepsilon _{r}(x)$ \emph{for }$x\in \left[ 0,\tau ^{-1}\left( T/2\right) %
\right] $\emph{\ satisfying above conditions. }

\subsection{Performance of GLK for computationally simulated data}

\label{sec:4.2}

In this subsection we present results of our computations for GLK for
computationally simulated data. For our testing we choose the coefficient $%
\varepsilon _{r}\left( x\right) $ whose graph is depicted on Fig. \ref{fig_4}%
-a). This is the same function as the one for AGCM. We use the FDM to solve
the forward problem (\ref{4.1})-(\ref{4.3}).\ The solution of this problem
generates the data $f(t):=u(0,t)$, $t\in \lbrack 0,\widetilde{T}],$ where $%
\widetilde{T}>0$ is a certain number. To verify the stability of our
computations with respect to the noise in the data, we introduce the random
noise in the data as in (\ref{2.37}), where $g\left( t\right) $ is replaced
with $f\left( t\right) $ and $T$ is replaced with $\widetilde{T}.$

Next, we use the change of variables $x\Leftrightarrow z$ in (\ref{4.6}) and
define the time interval $t\in (0,T),T=T(\widetilde{T}).$ Indeed, since in (%
\ref{4.15}) $z\in \lbrack 0,T/2]$ and $t\in \left[ -z,z\right] ,$ then the
value of $z_{\max }=T/2$ should be such that $\tau ^{-1}\left( z_{\max
}\right) :=x_{\max }\geq 1.$ Note that all what we need is a lower estimate
of $z_{\max }$ rather than its exact value. Hence, to get such an estimate,
it is sufficient to know a lower estimate for the unknown coefficient $%
\varepsilon _{r}\left( x\right) $ rather than the function $\varepsilon
_{r}\left( x\right) $ itself. Let 
\begin{equation*}
\left\{ z_{n}\right\} _{n=0}^{M_{z}}\subset \lbrack
0,T/2],0=z_{0}<z_{1}<...<z_{M_{z}}=T/2
\end{equation*}
be a partition of the interval $z\in \lbrack 0,T/2]$ into $M_{z}$ small
subintervals with the grid step size $h_{z}=z_{n}-z_{n-1}.$ For each $z_{n}$
we approximate the integral (\ref{4.15}) via the trapezoidal rule. For the
variable $t\in \left[ -z_{n},z_{n}\right] $ we use discrete values $\left\{
t_{k}=\pm z_{k}\right\} _{k=-n}^{k=n}.$ Next, we solve the resulting linear
algebraic system for each $z_{n}$ and find the discrete function $%
w(z_{n},t_{k}),\forall k\in \left[ -n,n\right] ,\forall n\in \left[ 0,M_{z}%
\right] .$ Next, we find the discrete function $\varepsilon _{r}\left( \tau
^{-1}\left( z_{n}\right) \right) $ via (\ref{4.17}). Finally, we numerically
invert (\ref{4.6}) and find the discrete function $\widetilde{\varepsilon }%
_{r}\left( x_{n}\right) $ for a certain partition $\left\{ x_{n}\right\}
_{n=0}^{M_{z}}$ of the interval $\left[ 0,x_{\max }\right] \supseteq \left[
0,1\right] .$ This discrete function $\widetilde{\varepsilon }_{r}\left(
x_{n}\right) $ is an approximate solution of CIP2.

Since the differentiation of noisy data is an unstable procedure and $\hat{f}%
^{\prime }(t-\lambda )$ is the kernel of the integral equation (\ref{4.15}),
we calculate the derivative $\hat{f}^{\prime }(t)$ in a special way. For
each point $t_{k}:=z_{k},k=1,...,M_{z}-1$ we approximate the derivative $%
\hat{f}^{\prime }(t_{k})$ as 
\begin{equation}
\hat{f}^{\prime }(t_{k})\approx \frac{\hat{f}\left( t_{k+1}\right) -\hat{f}%
\left( t_{k-1}\right) }{2h_{z}}.  \label{4.18}
\end{equation}%
Next, we extend this approximation as the discrete even function for $%
t_{k}=-z_{k},k=1,...,M-1.$ Even though (\ref{4.18}) seems to be a
non-regularizing procedure, it works quite well for our goal, since in the
discrete integration in (\ref{4.15}) the right hand side of (\ref{4.18}) is
actually multiplied by $h_{z}/2.$\newline

\begin{figure}[tb!]
\centering
\includegraphics[bb= 0 0 1340 1520, scale=0.25]{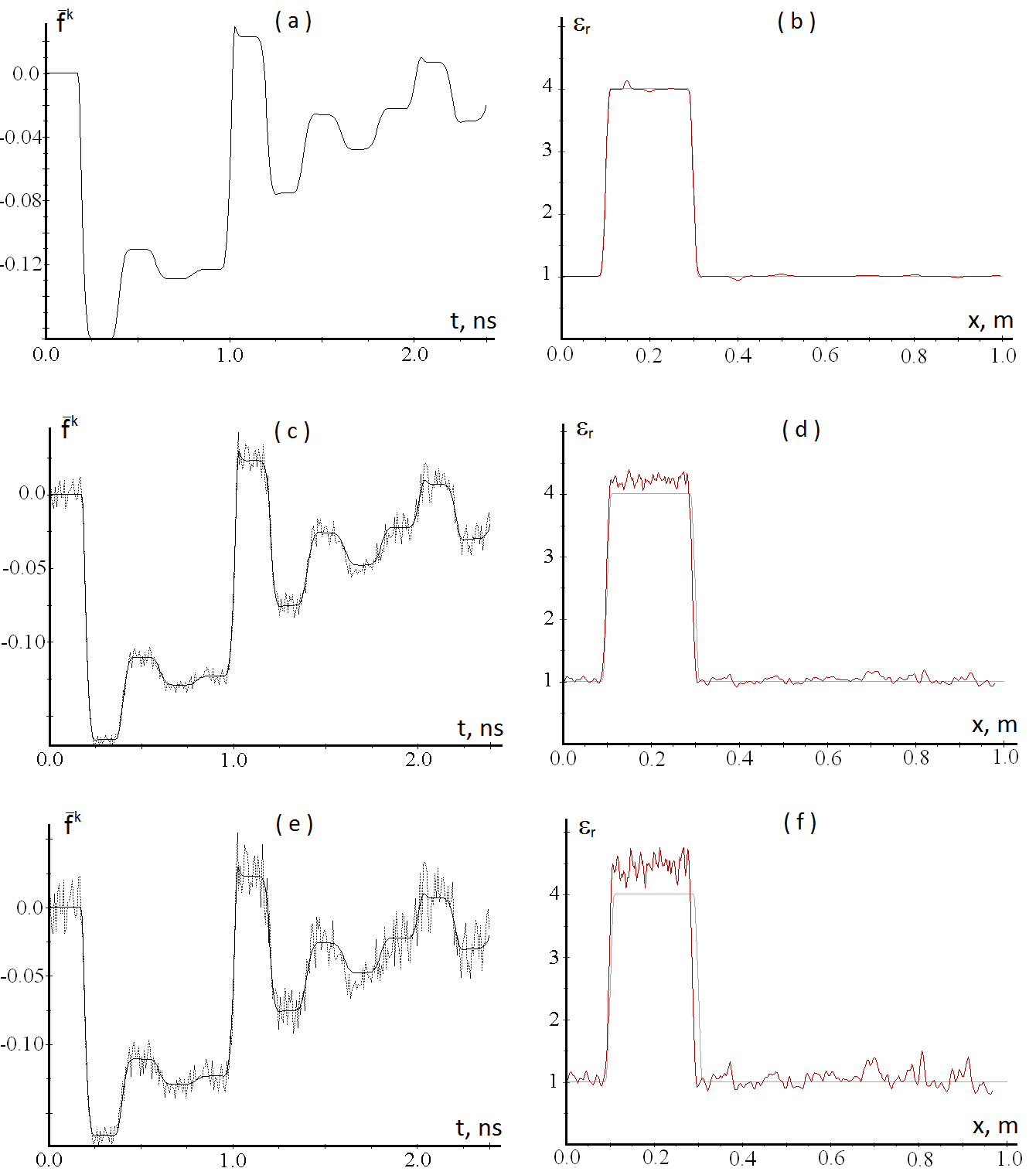}
\caption{ \textsf{\emph{Performance of GLK for the computationally simulated
data for the target of Fig.~\protect\ref{fig_4}-a). a), c) and e) the
computed function} $\overline{f}(t)$ \emph{in (\protect\ref{4.19}) for 0\%,
5\% and 10\% noise respectively. b), d) and f) computed function} $\protect%
\varepsilon_{r}(x)$ \emph{for cases a), c) and e) respectively.} }}
\label{fig_8}
\end{figure}

We test GLK for 0\%, 5\% and 10\% noise in the data $f(t)$.\ Let $M_{x}$ and 
$h_{x}$ be the number of grid points and the mesh size respectively for the
variable $x\in (0,1)$. We took $M_{x}=250$, which led to $h_{x}=0.004$.
After the change of variables (\ref{4.6}) we got $z\in \lbrack 0,1.168]$, $%
M_{z}=349$, $h_{z}=3.34\cdot 10^{-3}$. Results of computations are displayed
on Fig. \ref{fig_8}-a)-f). Note that Figures \ref{fig_8}-a), c), e) depict
the function 
\begin{equation}
\overline{f}(t)=-\frac{1}{4}(f(t)+1).  \label{4.19}
\end{equation}

Comparison of Figures \ref{fig_8}-b), d), f) with Figures \ref{fig_6}-a),
b), c) shows that GLK better images the shape of the target than AGCM. The
inclusion/background contrast $\max \varepsilon _{r}\left( x\right) $ is
computed accurately by both GLK and AGCM. On the other hand, the contrast
computed by AGCM does not change when the level of noise changes, while it
does change for GLK.

\section{Comparison of AGCM and GLK on Real Data}

\label{sec:5}

As it was pointed out in subsection 2.3.2, to work with experimental data,
it is necessary to figure out the calibration factor. This factor was
established for AGCM as $CF=10^{-7}$, see (\ref{602}). Since dielectric
constants of targets were not measured in experiments, while $CF=10^{-7}$
still led to acceptable values of those constants, then we believe that the
best is to use computed values of $\overline{R}$ in Table 1 to figure out
the calibration factor for GLK. Thus, we proceed as described in Subsection
5.1.

\subsection{The choice of the calibration factor for GLK}

\label{sec:5.1}

The procedure of this choice consists of the following three steps:

\emph{Step 1}. Select any target $X$ \ listed in Table 1. We call $X$
\textquotedblleft calibration target". Let $F_{X}\left( t\right) $ be the
data for the target $X,$ which are pre-processed as in subsection 2.3.1, see
Fig. \ref{fig_5}.

\emph{Step 2}. Let $R_{GLK}\left( X\right) $ and $R_{AGCM}\left( X\right) $
be the function $R\left( x\right) $ in (\ref{1}) computed by GLK and AGCM
respectively for the target $X$. Let $\overline{R}_{GLK}\left( X\right) $
and $\overline{R}_{AGCM}\left( X\right) $ be corresponding values of the
number $\overline{R}$ in (\ref{2}). Hence, $\overline{R}_{AGCM}\left(
X\right) $ is the number $\overline{R}$ for the target $X$ in Table 1. Let $%
CF\left( X\right) $ be the calibration factor for GLK. Thus, GLK should be
computed for the data $CF\left( X\right) \cdot F_{X}\left( t\right) .$
Choose such a calibration factor $CF\left( X\right) $ that $\overline{R}%
_{GLK}\left( X\right) =\overline{R}_{AGCM}\left( X\right) .$ Thus, $%
\overline{R}_{GLK}\left( X\right) $ is computed via GLK for the data $%
CF\left( X\right) \cdot F_{X}\left( t\right) $ and $\overline{R}%
_{AGCM}\left( X\right) $ is computed via AGCM for the data $%
10^{-7}F_{X}\left( t\right) ,$ see (\ref{602}).

\emph{Step 3}. Let $Y\neq X$ be any other target listed in Table 1. When
applying GLK to the data $F_{Y}\left( t\right) $, use the same calibration
factor $CF\left( X\right) .$

\subsection{Results}

\label{sec:5.2}

When applying GLK to the pre-processed data, we have chosen $T$ to be the
moment of time which is 1.33 nanoseconds to the right from the end of the
selected peak.

\begin{figure}[tb!]
\centering
\includegraphics[bb= 0 0 1360 420, scale=0.3]{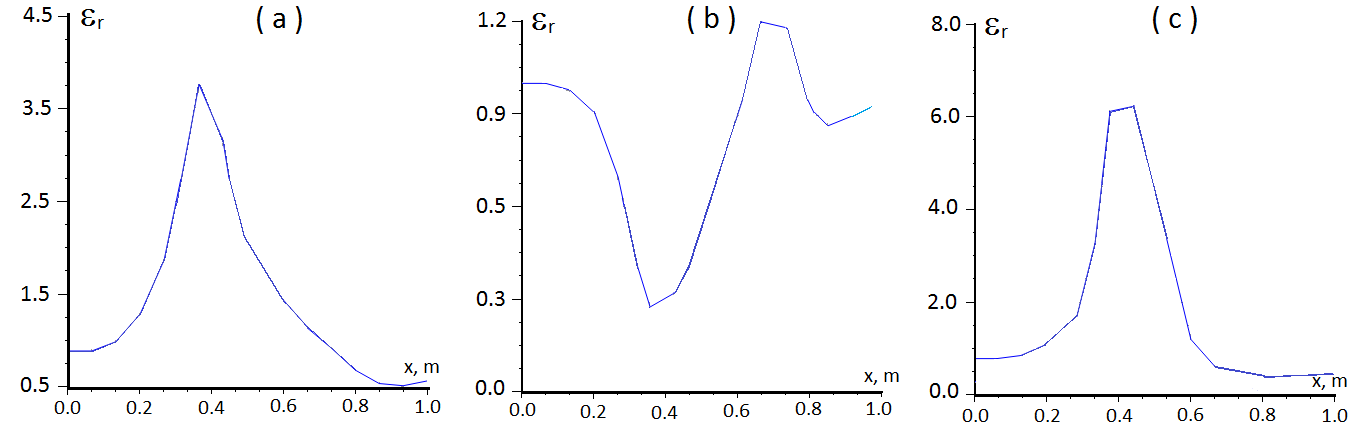}
\caption{ \textsf{\ \emph{a) The function} $R_{GLK}(x)$ \emph{\ for the wood
stake when it was the calibration target.\ Here} $CF(\text{wood stake}%
)=10^{-5}$, $\overline{R}_{GLK}=3.8$; b) \textsf{\ \emph{the function} $%
R_{GLK}(\text{plastic cylinder}) $\emph{\ for the case when the plastic
cylinder was the calibration target.\ Here} $CF(\text{plastic cylinder}%
)=1.8\cdot 10^{-5}$, $\overline{R}_{GLK}(\text{plastic cylinder})=0.28$; c)
the function } $R_{GLK}(\text{bush})$ \emph{\ for the case when bush was the
calibration target.\ Here} $CF(\text{bush})=0.6\cdot 10^{-5}$, $\overline{R}%
_{GLK}(\text{bush})=6.4$.}}
\label{fig_9}
\end{figure}

\textbf{Case 1.} Wood stake is the calibration target. Using Step 2 of
Subsection 5.1, we have obtained $CF\left( \text{wood stake}\right) =10^{-5}.
$ Figure~\ref{fig_9}-a displays the function $R_{GLK}\left( \text{wood stake}%
\right) $.

\begin{center}
\begin{table}[h!]
\caption{Case 1. Computed numbers $\overline{R}_{GLK}$ for four other
targets in comparison with the values of $\overline{R}$ taken from Table~1.
Here wood stake was the calibration target. Bold faced is an unrealistic
value of $\overline{R}_{GLK}$. }%
\begin{tabular}{|l|c|c|}
\hline
Target & $\overline{R}_{GLK}$ & $\overline{R}$ from Table 1 \\ \hline
metal cylinder & 3.25 & 4.3 \\ \hline
metal box & 3.97 & 3.8 \\ \hline
plastic cylinder & 0.6 & 0.28 \\ \hline
bush & \textbf{42.5} & 6.4 \\ \hline
\end{tabular}%
\end{table}
\end{center}

We observe from Table 2 that the value $\overline{R}_{GLK}\left( \text{bush}%
\right) =42.5$ is significantly overestimated compared with the published
interval $\overline{R}\in \lbrack 3,20]$ in \cite{vegetation}.

\textbf{Case 2}. Plastic cylinder is the calibration target. We have
obtained $CF(\text{plastic cylinder})=1.8\cdot 10^{-5}$. Figure~\ref{fig_9}%
-b displays the function $R_{GLK}(\text{plastic cylinder})$ with this
calibration factor.

\begin{center}
\begin{table}[h!]
\caption{Case 2. Computed numbers $\overline{R}_{GLK}$ for four other
targets in comparison with values of $\overline{R}$ taken from Table~1. Here
plastic cylinder was the calibration target. Bold faced are those values of $%
\overline{R}_{GLK}$ which result in unrealistic values of dielectric
constants.}%
\begin{tabular}{|l|c|c|}
\hline
Target & $\overline{R}_{GLK}$ & $\overline{R}$ from Table 1 \\ \hline
metal cylinder & 9.8 & 4.3 \\ \hline
metal box & 13.9 & 3.8 \\ \hline
wood stake & \textbf{17} & 3.8 \\ \hline
bush & $\mathbf{>>}$\textbf{100} & 6.4 \\ \hline
\end{tabular}%
\end{table}
\end{center}

We observe from Table 3 that the values of $\overline{R}_{GLK}$ for the wood
stake and bush are significantly overestimated, compared with the published
intervals $[2,6]$ of \cite{Table1} and $[3,20]$ of \cite{vegetation}
respectively. Actually we had a blow in the image of bush via GLK in this
case.

\textbf{Case 3.} Bush is the calibration target. We have obtained $CF(\text{%
bush})=0.6\cdot 10^{-5}.$ Figure~\ref{fig_9}-c) displays the function $%
R_{GLK}(\text{bush})$ with this calibration factor.

\begin{center}
\begin{table}[h!]
\caption{Case 3. Computed numbers $\overline{R}_{GLK}$ for four other
targets in comparison with values of $\overline{R}$ taken from Table~1. Here
bush was the calibration target. Bold faced are those values of $\overline{R}%
_{GLK}$ which lead to unrealistic values of dielectric constants.}%
\begin{tabular}{|l|c|c|}
\hline
Target & $\overline{R}_{GLK}$ & $\overline{R}$ from Table 1 \\ \hline
metal cylinder & 2.0 & 4.3 \\ \hline
metal box & \textbf{1.16} & 3.8 \\ \hline
wood stake & 2.2 & 3.8 \\ \hline
plastic cylinder & \textbf{0.74} & 0.28 \\ \hline
\end{tabular}%
\end{table}
\end{center}

It follows from Table~4 that the value of $\overline{R}_{GLK}$ for metal box
is significantly underestimated.\ Indeed, since in the dry sand, which was
the background here, $\varepsilon _{r}\in \left[ 3,5\right] $ \cite{Table1},
then, using (\ref{2}), we obtain $\varepsilon _{r}\left( \text{metal box}%
\right) \in \left[ 3.48,5.8\right] $. On the other hand, by (\ref{2.38}) we
should have $\varepsilon _{r}\left( \text{metal box}\right) \in \left[ 10,30%
\right] $. Thus, we have the underestimation of $\varepsilon _{r}\left( 
\text{metal box}\right) $ by the factor of 1.72. In addition, the value of $%
\overline{R}_{GLK}$ for plastic cylinder is overestimated in Table 4.
Indeed, since $\varepsilon _{r}\left( \text{background}\right) \in \left[ 3,5%
\right] ,$ then it follows from this table that $\varepsilon _{r}\left( 
\text{plastic cylinder}\right) \in \left[ 2.22,3.7\right] $. On the other
hand, Table \cite{Table2} tells us that we should have $\varepsilon
_{r}\left( \text{plastic cylinder}\right) \approx 1.2.$ Hence, we have
overestimation of $\varepsilon _{r}\left( \text{plastic cylinder}\right) $
by the factor of 1.85.

Our calculations for the case when metal box was chosen as calibration
target led to the same results as those in Table 2. This is because $%
\overline{R}_{AGCM}\left( \text{wood stake}\right) =\overline{R}%
_{AGCM}\left( \text{metal box}\right) =3.8.$ Also, $\overline{R}%
_{AGCM}\left( \text{metal cylinder}\right) =4.3\approx \overline{R}%
_{AGCM}\left( \text{metal box}\right) .$ Hence, the calculation for the case
when the metal cylinder was chosen as calibration target again led to
results similar with those of Table 2.

\begin{figure}[tb!]
\centering
\includegraphics[bb= 0 0 1330 470, scale=0.3]{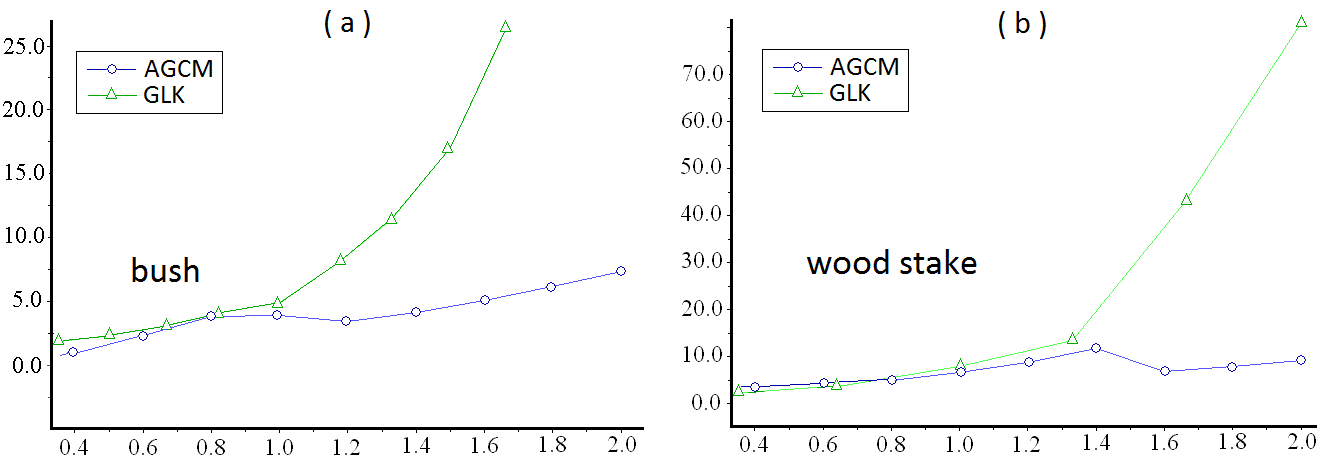}
\caption{ \textsf{\ \emph{Dependencies of} $\overline{R}_{AGCM}$ \emph{\ and}
$\overline{R}_{GLK}$ \emph{\ from the calibration factor for bush and wood
stake are displayed. In both a), b) ``1'' for AGCM corresponds to} $%
CF=10^{-7}$, \emph{\ see (\protect\ref{602}). a) ``1'' corresponds} $CF(%
\text{bush})=0.6\cdot 10^{-5}$ \emph{\ (Case 3). b) ``1'' corresponds} $CF(%
\text{wood stake})=10^{-5}$. }}
\label{fig_10}
\end{figure}

To see how the changes in the calibration factor affect the reconstructed
values of $\overline{R}_{AGCM}$ and $\overline{R}_{GLK},$ we present Figures %
\ref{fig_10}. The visual analysis of these curves shows that the increase of
the calibration factor affects $\overline{R}_{AGCM}$ linearly, and it
affects $\overline{R}_{GLK}$ exponentially. This explains results of above
Cases 1-3.

\section{Discussion}

\label{sec:6}

Theorem 2 guarantees existence and uniqueness of the solution of the GLK
equation only when the data $f(t)$ belongs to the range of the operator of
the forward problem, i.e. for the case of errorless data. However, in the
realistic case of an error in the data, stability with respect to this error
is not guaranteed by Theorem 2. Hence, the question about regularizing
properties of GLK remains open. On the other hand, estimate (\ref{2.36}) of
Theorem 1 ensures stability of AGCM with respect to a small error in the
data, and the same is true for the 3-d version of this method \cite%
{BK,BK3,Kuzh}.

We have used both computationally simulated and experimental data to compare
numerical performances of two methods: GLK and AGCM. In the case of
computationally simulated data, the above difference between Theorems 1 and
2 is slightly reflected in the different behavior of computed
inclusion/background contrasts in the presence of noise: these contrasts
were not changing with noise in the data for AGCM and were slightly changing
for GLK.\ On the other hand, in the case of the above synthetic data, GLK
provides better images of shapes of targets, compared with AGCM.

As it was pointed out in subsection 2.3.2, to have unbiased studies of these
experimental data, it is important that the calibration factor should be
chosen the same for all five targets of Table 1. However, it was shown in
subsection 5.2 that it is impossible to choose such a calibration factor for
GLK, which would provide satisfactory values of inclusion/background
contrasts for all five targets. On the other hand, AGCM has worked
successfully with the uniform calibration factor $CF=10^{-7}$ for all five
targets in the case of blind data. We point out that these conclusions are
relevant only for the specific case of the above set of experimental data
and for the above specific data pre-preprocessing procedure. On the other
hand, we do not possess other sets of 1-d experimental data at the moment,
and we are also unaware about other appropriate data pre-processing
procedures.

To explain Fig. \ref{fig_10}, we present the GLK equation (\ref{4.15})\emph{%
\ }as 
\begin{equation}
w(z,t)-\frac{CF\left( X\right) }{2}\int\limits_{-z}^{z}\hat{f}^{\prime
}(t-\tau )w(z,\tau )d\tau =\frac{1}{2},\ \ \ t\in \lbrack -z,z],\ \ \forall
z\in \lbrack 0,T/2].  \label{6.1}
\end{equation}%
In (\ref{6.1}) $f(t)$ is the pre-preprocessed data as in in subsection 2.3.1
and $\hat{f}(t)$ is the odd extension of $f(t)$. Denote $\beta :=z\cdot
CF\left( X\right) \sup_{\left\vert t\right\vert <T/2}\left\vert \hat{f}%
^{\prime }(t)\right\vert .$ If $\beta \in \left( 0,1\right) $, then one can
solve integral equation (\ref{6.1}) via the resolvent series, see, e.g. the
book \cite{V}. It is clear from this series that the solution $w(z,t)$
changes almost with an exponential speed when the calibration factor $%
CF\left( X\right) $ changes. This explains, at least partially, the
exponential behavior of GLK curves of Fig.~\ref{fig_10}. This is because the
data for the inverse problem are introduced in the operator. 

On the other hand, almost linear behavior of AGCM curves on Fig. \ref{fig_10}
can be explained by the fact that AGCM uses the logarithm of the solution $w$
of the problem (\ref{2.8}), (\ref{2.9}), see (\ref{2.120}).

\begin{center}
\textbf{Acknowledgments}
\end{center}

This research was supported by US Army Research Laboratory and US Army
Research Office grant W911NF-11-1-0399, Integration project number 14 of
Siberian Branch of Russian Academy of Science (SB RAS), Collaboration
project number~12-2013 between SB RAS and NAS of Ukraine, and grant
12-01-00773 of Russian Foundation of Basic Research.

\end{document}